\documentclass[rnote]{aa} 
\usepackage{graphicx}
\usepackage{txfonts}
\begin{document}
   \title{Microlensing towards the LMC revisited by adopting a
non--Gaussian velocity distribution for the sources}


   \author{L. Mancini
          \inst{1,2}
          }

   \institute{Dipartimento di Fisica ``E.R. Caianiello'',
    Universit$\mathrm{\grave{a}}$ di Salerno, via S. Allende, Baronissi (SA), Italy\\
        \email{lmancini@physics.unisa.it}
         \and
             Istituto Nazionale di Fisica Nucleare, Sezione di
                Napoli, Italy\\
             }

   \date{}


  \abstract
   {}
   {We discuss whether the Gaussian is a reasonable approximation of the
velocity distribution of stellar systems that are not spherically
distributed.}
   {By using a non-Gaussian
velocity distribution to describe the sources in the Large
Magellanic Cloud (LMC), we reinvestigate the expected microlensing
parameters of a lens population isotropically distributed either
in the Milky Way halo or in the LMC (self lensing). We compare our
estimates with the experimental results of the MACHO
collaboration.}
   {An interesting result that
emerges from our analysis is that, moving from the Gaussian to the
non-Gaussian case, we do not observe any change in the form of the
distribution curves describing the rate of microlensing events for
lenses in the Galactic halo. The corresponding expected timescales
and number of expected events also do not vary. Conversely, with
respect to the self-lensing case, we observe a moderate increase
in the rate and number of expected events. We conclude that the
error in the estimate of the most likely value for the MACHO mass
and the Galactic halo fraction in form of MACHOs, calculated with
a Gaussian velocity distribution for the LMC sources, is not
higher than $2\%$.}
   {}

   \maketitle
%

\section{Introduction}

Galaxies are complex, collisionless, gravitationally-bound systems
formed by secular gravitational self-interaction and collapse of
its constituents. Significant progress has been made in both
observational and theoretical studies developed to improve our
understanding of the evolutionary history of galaxies and the
physical processes driving their evolution, leading to the Hubble
sequence of galaxy type that we observe today. However, many
aspects of their features, such as morphology, compositions, and
kinematics, still remain unclear. In particular, it is not obvious
how the velocities of their constituent (in particular stellar)
components can be described, because we cannot consider them to be
isotropically distributed at any point. Little is known about the
velocity distribution (VD) of the stellar populations of galactic
components. While the distribution of stellar velocities in an
elliptical galaxy is generally reasonably close to a Gaussian,
analyses of the line-of-sight (l.o.s.) velocity distributions of
disk galaxies have shown that these distribution are highly
non-Gaussian (\cite{binney98}).

Today, one of the most important problems regarding the
composition of the Milky Way (MW) concerns the existence of dark,
compact agglomerates of baryons in the Galactic halo, the
so-called MACHOs (MAssive Compact Halo Objects). From the
experimental point of view, several observational groups have
attempted to detect these objects by performing microlensing
surveys in the directions of the Large Magellanic Cloud (LMC),
Small Magellanic Cloud, and M31. Two groups (MACHO and
POINT-AGAPE) reported similar conclusions, despite the fact that
they observed different targets (LMC and M31), that is roughly
$20\%$ of the halo mass must be in the form of MACHOs (Alcock et
al. \cite{alcock00}; \cite{calchi05}). However, the interpretation
of their data is controversial because of the insufficient number
of events detected, and the existing degeneration among the
parameters. Discordant results have been reported by other
experimental teams (\cite{tisserand07}; \cite{dejong04}).

Accurate theoretical estimates of the microlensing parameters,
supported by statistical analysis, are fundamental to the
interpretation of the experimental results. However, there are
many uncertain assumptions in the adopted lens models. These
uncertainties, that could lead towards an incorrect interpretation
of the data, are mostly related to the shape of the individual
galactic components and the kinematics of the lens population.

One of the first problems to be raised  by the scientific
community concerned the shape of the Galactic dark halo.
Unfortunately, information that can be extracted from observations
of high-velocity stars and satellite galaxies does not place
strong constraints on its shape. In the absence of precise data,
we are aided by computational models of the formation of galaxies,
which suggest that the dark halos are more or less spherical
(\cite{navarro96}). However, Griest (\cite{Griest91}) showed that,
referring to MACHOs, the optical depth is relatively independent
of assumptions about the core and cutoff radii of the MW halo.
Sackett $\&$ Gould \cite{sackett93} first investigated the role of
the MW halo shape in Magellanic Cloud lensing, finding that the
ratio of the optical depths towards the Small and Large Magellanic
Clouds was an indicator of the flattening of the Galactic dark
halo. Alcock et al. (\cite{alcock00}) considered a wide family of
halos, besides the spherical one, ranging from a massive halo with
a rising rotation curve to models with more massive disks and
lighter halos. On the other hand, the problem related to the shape
of the LMC halo was defined by Mancini et al. (\cite{mancini04}).
These authors also explored the consequences of different LMC
disk/bar geometries a part from the coplanar configuration. All
these studies demonstrated that the estimate of the microlensing
parameters were noticeably affected by the shape of the Galactic
halo and the other Galactic components.

For the kinematics of the lenses, the expression of their
random-motion velocity was reanalyzed by Calchi Novati et al.
(\cite{calchi06}), who considered the LMC bulk motions including
the drift velocity of the disk stars. This study indicates that
the mean rotational velocity of the LMC stars is irrelevant to
estimates of the MACHO microlensing parameters due to the
preponderance of the bulk motion of the LMC. For the self-lensing,
it is slightly significant for lenses located in the bar and
sources in the disk. We emphasize that the VD of the LMC sources
has always been modeled by a Gaussian. This assumption is just a
first approximation and was adopted for practical reasons. In this
paper, we investigate whether this hypothesis is acceptable or not
for different source/lens configurations or at least provide a
quantitative measure of the effectiveness and accuracy of the
Gaussian hypothesis. To achieve this purpose, we re-examine the
framework of microlensing towards the LMC, and in particular we
recalculate the number of expected events by assuming that the
source velocities are no longer Gaussian distributed. Both the
MACHO and the self-lensing cases are considered.

\section{Non-Gaussian velocity distributions}\label{nonGaussian_VD}
If we consider a spherically symmetric distribution of stars with
density $\rho$, then we can describe the dynamical state of the
system by a distribution function of the following form
\begin{equation}
F \left(E \right)= \frac{\rho}{\left( 2\pi\sigma^2 \right)^{3/2}}
~ e^{E / \sigma^{2}},
 \label{Eq_DF}
\end{equation}
where $E=\Psi-v^{2}/2$ is the binding energy per unit mass, and
$\Psi$ is the relative gravitational potential (\cite{binney87}).
It is well known that the structure of a collisionless system of
stars, whose density in phase space is given by Eq. (\ref{Eq_DF}),
is identical to the structure of an isothermal self-gravitating
sphere of gas. Therefore, the velocity distribution at each point
in the stellar-dynamical isothermal sphere is the Maxwellian
distribution $f(v)=N e^{-\frac{1}{2}v^2/\sigma^2}$, which equals
the equilibrium Maxwell-Boltzmann distribution given by the
kinetic theory. If we consider a stellar system that is far from
having a spherical distribution (for example, a galactic flattened
disk, a triaxial bulge, or an elongated bar), we do not expect
that it is correct to use a Maxwellian distribution to describe
its velocity profile. In the same way, we must ask if it is
correct or not to use a Gaussian shape
$f(v)\sim\mathrm{exp}(-(v^2/\sigma^2))$ to describe the l.o.s. or
projected velocity profiles of non-spheroidal galactic components.
We attempt to answer this question by using non-Gaussian VDs
obtained by simulations. Numerical simulations of collapse and
relaxation processes of self-gravitating collisionless systems,
similar to galaxies, are useful in identifying their general
trends, such as density and anisotropy profiles. Two studies of
the velocity distribution function of these systems by means of
numerical simulations were performed, showing that the velocity
distribution of the resultant quasi-stationary states generally
becomes non-Gaussian (Iguchi et al. \cite{iguchi05}; Hansen et al.
\cite{hansen06}).

\subsection{Superposition of Gaussian distributions}
N-body simulations of different processes of galaxy formation were
performed by Iguchi et al. (\cite{iguchi05}). As a result of their
simulations, these authors found stationary states characterized
by a velocity distribution that is well described by an equally
weighed superposition of Gaussian distributions of various
temperatures, a so-called {\it democratic temperature
distribution} (DT distribution), that is
\begin{equation}
f_{\mathrm{DT}}\left(v\right)=\frac{1}{\sigma^2} \left\{
\sqrt{\frac{2}{\pi}} \sigma e^{-v^2/(2 \sigma^2)} -|v| \left[
1-\mathrm{Erf}\left( \frac{|v|}{\sqrt{2} \sigma} \right) \right]
\right\}, \label{Eq_DTVD}
\end{equation}
where $\mathrm{Erf}(x)$ is the error function. The conclusion is
that the DT velocity distribution is a universal property of
self-gravitating structures that undergo violent, gravitational
mixing. The origin of such universality remains, however, unclear.

\subsection{Universal velocity distribution}
%
Hansen et al. (\cite{hansen06}) performed a set of simulations of
{\it controlled collision} experiments of individual purely
collisionless systems formed by self-gravitating particles. They
considered structures initially isotropic as well as highly
anisotropic. After a strong perturbation followed by a relaxation,
the final structures were not at all spherical or isotropic. The
VD extracted from the results of the simulations was divided into
radial and tangential parts. In this way, they found that the
radial and tangential VDs are universal since they depend only on
the radial or tangential dispersion and the local slope of the
density; the density slope $\alpha$ is defined as the radial
derivative of the density $\alpha\equiv\mathrm{d} \ln\rho
/\mathrm{d}\ln r$.
The points obtained by the simulations, which describe the
universal tangential VD, are described well by the following
functional form,
\begin{equation}
f_{\mathrm{tan}}\left(v_{\mathrm{tan}}\right)= \frac{
v_{\mathrm{tan}}}{k^{2} \pi \sigma_{\mathrm{tan}}^{2}} \left( 1-
\left(1-q\right)
\left(\frac{v_{\mathrm{tan}}}{k\,\sigma_{\mathrm{tan}}}\right)^2
\right)^{q/(1-q)} \label{Eq_TangentialVD},
\end{equation}
where $\sigma_{\mathrm{tan}}$ is the tangential velocity
dispersion, $v_{\mathrm{tan}}$ is the two-dimensional velocity
component projected on the plane orthogonal to the l.o.s., while
$q$ and $k$ are free parameters. Hansen et al. (\cite{hansen06})
reported the universal tangential VD for three different values of
the density slope $\alpha$. Here we use the intermediate case
where $\alpha$ equals -2. This VD has a characteristic break,
where $v_{\mathrm{tan}}=1.6 \, \sigma_{\mathrm{tan}}$ is taken to
be the transition velocity. The low energy part is described by
$q=5/3$ and $k=0.93$. By comparison, for the high energy tails,
the parameters are $q=0.82$ and $k=1.3$.


\section{Microlensing towards the LMC revisited}
%
Concerning the Hubble sequence type, the NASA Extragalactic
Database considers the LMC as Irr/SB(s)m. The LMC is formed of a
disk and a prominent bar at its center, suggesting that it may be
considered as a small, barred, spiral galaxy. Different
observational campaigns towards the LMC (MACHO, EROS, OGLE, MOA,
SUPERMACHO) have been performed with the aim of detecting MACHOs.
Among these, only the MACHO and EROS groups have published their
results. The EROS collaboration detected no events
(\cite{tisserand07}). In contrast, the MACHO Project detected 16
microlensing events, and concluded that MACHOs represent a
substantial part of the Galactic halo mass, but is not the
dominant component (Alcock et al. \cite{alcock00}). The maximum
likelihood estimate of the mass $m$ of the lensing objects was
$\approx 0.5$ M$_{\odot}$, whereas the fraction $f$ of dark matter
in the form of MACHOs in the Galactic halo was estimated to be
$\sim 20\%$.

In the numerical estimates of the microlensing parameters, useful
in studying the fraction of the Galactic halo in the form of
MACHOs, a Gaussian shape velocity distribution is still commonly
used to describe the projected velocity distribution for the
lenses as well as the source stars, although they are not
spherically distributed (\cite{jetzer02}; Mancini et al.
\cite{mancini04}; \cite{assef06}; Calchi Novati et al.
\cite{calchi06}). Here, our intention was to utilize the
non-Gaussian velocity profiles described in the previous section
for the sources, instead of the usual Gaussian shape, and show how
the microlensing probabilities change accordingly. As a concrete
case, in Sect. 3.1 we analyzed two main parameters of the
microlensing towards the LMC, the rate and the number of expected
microlensing events generated by a lens population belonging to
the MW halo as well as one belonging to the LMC itself. The
results of our model were compared with the MACHO collaboration
observational results (Alcock et al. \cite{alcock00}). Finally,
the method of maximum likelihood is used in Sect. 3.2 to calculate
the probability isocontours in the $\{m,f\}$ plane.

%
\subsection{Microlensing rate and number of expected events}
\begin{figure}
   \centering
 \resizebox{8cm}{!}{\includegraphics{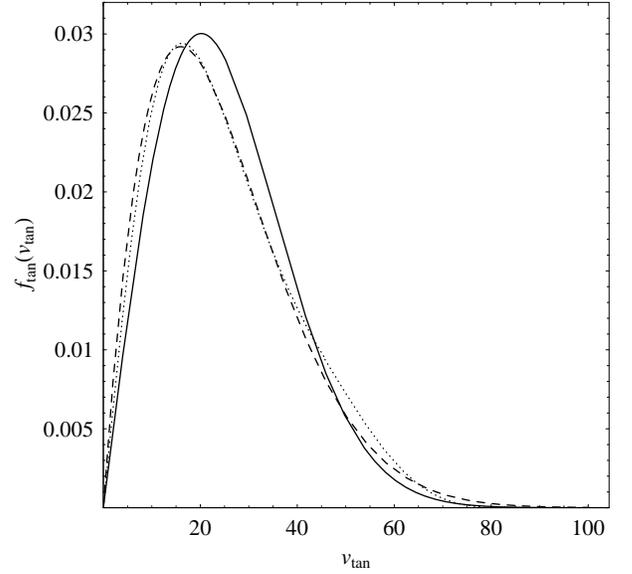}}
 \caption{The velocity profiles used to describe the kinematics
 of the LMC stars: VDs derived from the simulations of
 Iguchi et al. (\cite{iguchi05}) (dashed curve) and
 Hansen et al. (\cite{hansen06}) (dotted line).
 The solid black line represents a Gaussian profile. }
 \label{Fig VDP1}
\end{figure}

%
We restricted our analysis by considering a homogeneous subset of
12 Paczy\'{n}ski-like events taken from the original larger set B
reported by MACHO; we did not consider the Galactic disk events,
the binary event, and all candidates whose microlensing origin had
been placed in doubt. In our calculations, we used the models
presented in Mancini et al. (\cite{mancini04}) to represent the
various Galactic components: essentially an isothermal sphere for
the Galactic halo, a sech$^2$ profile for the LMC disk, and a
triaxial boxy-shape for the LMC bar. \cite{marel02} measured the
velocity dispersion of the LMC source stars to be $20.2$ km/s.
This measurement was completed as usual by a quantitative analysis
of the absorption lines in the LMC spectrum, by assuming a
Gaussian form for the VD. In principle, to obtain an estimate of
the velocity dispersion for a non-Gaussian distribution, we have
to repeat the same analysis of the LMC line profile by applying a
non-Gaussian algorithm. To a first approximation, we ignored this
subtlety and simply assumed that the dispersion of Gaussian and
non-Gaussian VDs were equal. Fixing in this way the values of the
velocity dispersions, we draw in Fig. \ref{Fig VDP1} the universal
tangential VD (dotted line) together with the TD distribution
(dashed curve), and a classical Gaussian profile (solid line).

In general, the velocity of the lenses $v_{\mathrm{\ell}}$
consists of a global rotation plus a dispersive component. Since
we assumed that the MW halo has a spherical form, we considered
that the lenses are spherically distributed. In this case, the
rotational component could be neglected, and at the same time we
could safely consider the distribution of the dispersive component
to be isotropic and Maxwellian (\cite{derujula95}). This
assumption was also supported by an analysis of the kinematics of
nearly 2500 Blue Horizontal-Branch Halo stars at $|z|\geq4$ kpc,
and with distances from the Galactic center up to $\sim60$ kpc
extracted from the Sloan Digital Sky Survey, where the observed
distribution of l.o.s. velocities is well-fitted by a Gaussian
distribution (\cite{xue08}).

It is well known that the number of events $N$ is the sum, $N=\sum
N_{\mathrm{field}}$, of the number of events expected for each
monitored field of the experiment defined to be
$N_{\mathrm{field}}=E_{\mathrm{field}}
\int_{0}^{\infty}(\mathrm{d}\Gamma/\mathrm{d}T_{\mathrm{E}}) \,
{\cal{E}}(T_{\mathrm{E}})\,\mathrm{d} T_{\mathrm{E}}$,
%
%
where $E_{\mathrm{field}}$ is the field exposure,
$\mathrm{d}\Gamma/\mathrm{d}T_{\mathrm{E}}$ is the differential
rate with respect to the observed event duration, $T_{\mathrm{E}}$
is the Einstein time, and ${\cal{E}}(T_{\mathrm{E}})$ is the
detection efficiency of the experiment. The differential rate is
defined to be (Mancini et al. \cite{mancini04}; Calchi Novati et
al. \cite{calchi06})
\begin{eqnarray}
\frac{\mathrm{d}\Gamma}{\mathrm{d}T_{\mathrm{E}}}&=& %
\int_{0}^{2\pi} \mathrm{d}\alpha %
\int_{0}^{2 \pi}\mathrm{d}\varphi %
\int_{-\pi/2}^{\pi/2} \cos \theta  \, \mathrm{d}\theta %
\int_{0}^{\infty}
f\left(v_{\mathrm{s}}\right)\mathrm{d}v_{\mathrm{s}} \times\nonumber\\ %
&\times& %
\int_{0}^{\infty} \frac{v_{\mathrm{\ell}}^2}{2 \pi \, \sigma_{\mathrm{\ell}}^2} %
\exp\left(-\frac{v_{\mathrm{\ell}}^2+x^2v_{\mathrm{s}}^2+2x \,
v_{\mathrm{\ell}}v_{\mathrm{s}}\cos
\varphi}{2\sigma_{\mathrm{\ell}}}
\right)\mathrm{d}v_{\mathrm{\ell}}%
\times\\
&\times& %
\int_{\mu_{\mathrm{min}}}^{\mu_{\mathrm{max}}}
\frac{R_{\mathrm{E}}}{\cal N}
\frac{\mathrm{d}n(x)}{\mathrm{d}\mu} \, \mathrm{d}\mu%
\int_{0}^{1} x \, \mathrm{d}x %
\int_{d_{\mathrm{min}}}^{d_{\mathrm{min}}}
\rho_{\mathrm{s}}(D_{\mathrm{os}})
D_{\mathrm{os}}\,\mathrm{d}D_{\mathrm{os}} \nonumber, \label{Eq_rate}%
\end{eqnarray}
where $\rho_{\mathrm{s}}$ is the source density,
$f(v_{\mathrm{s}})$ represents the two-dimensional transverse
velocity distribution of the sources, $x$ is the ratio between the
observer-lens distance $D_{\mathrm{ol}}$ and the observer-source
distance $D_{\mathrm{os}}$, whereas $\mu$ is the lens mass in
solar mass units. The normalization factor $\cal N$ is the
integral over the l.o.s. of the sources.
The distribution $\mathrm{d}n(x)/\mathrm{d}\mu$ represents the
number of lenses with mass between $\mu$ and $\mu+\mathrm{d}\mu$
at a given point in the Galactic component considered. Assuming
the {\it factorization hypothesis}, we can write
$\mathrm{d}n(x)/\mathrm{d}\mu$ as the product of a distribution
$\mathrm{d}n_{0}/\mathrm{d}\mu$ depending only on $\mu$ and the
pertinent density profile (\cite{derujula95}; Mancini et al.
\cite{mancini04}) $\frac{\mathrm{d}n(x)}{\mathrm{d}\mu}=
\frac{\rho_{\mathrm{\ell}}(x)}{\mathrm{M}_{\sun}}
\frac{\mathrm{d}n_{0}}{\mathrm{d}\mu}$,
where $\rho_{\mathrm{\ell}}$ is the lens density. Concerning the
functional form of $\mathrm{d}n_{0}/\mathrm{d}\mu$, we supposed
that for the lenses in the halo the mass function is peaked at a
particular mass $\mu_{0}$, so that it could be described by a
delta function $\mathrm{d}n_{0}/\mathrm{d}\mu=
\delta(\mu-\mu_{0})/\mu_{0}$.
%
%
For lenses in the LMC disk/bar, we utilized the exponential form
$\mathrm{d}n_{0}/\mathrm{d}\mu=A\, \mu^{-\alpha} e^{-(
\mu_{0}/\mu)^{\beta}}$ (\cite{chabrier01}), where $\alpha=3.3$,
$\mu_{0}=716.4$, $\beta=0.25$, whereas $A$ is obtained from the
normalization condition $\int_{0.08}^{10}A\mu^{1-\alpha}e^{-
(\mu_{0}/\mu)^{\beta}} \mathrm{d}\mu=1$.

%
%
\subsubsection{Lenses in the Galactic halo}\label{rateandnev}
We calculated the differential rate of the microlensing events
with respect to the Einstein time, along the lines pointing
towards the events found by the MACHO collaboration in the LMC and
for different values of $\mu_{0}$. We used a Gaussian VD for
$f(v_{\mathrm{s}})$ as well as the non-Gaussian VDs, given by Eqs.
(\ref{Eq_DTVD}) and (\ref{Eq_TangentialVD}). As $\mu_{0}$ and the
l.o.s. change, we did not observe any substantial reduction in the
height of the distribution curve of the microlensing event rate,
and the corresponding expected timescale did not vary among the
cases considered. With respect to the number of events, the
situation did not change. Taking into account the MACHO detection
efficiency and the total exposure, we calculated the expected
number of events, summed over all fields examined by the MACHO
collaboration in the case of a halo consisting ($100\%$) of
MACHOs. Both in the Gaussian and the non-Gaussian case, we
achieved the well-known result that the expected number of events
is roughly 5 times higher than observed.

\subsubsection{Self-lensing}
We repeated the same analysis for the self-lensing configuration,
that is where both the lenses and the sources are located in the
disk/bar of the LMC.
%
%
By varying the l.o.s., we found in general that the microlensing
differential rate for the non-Gaussian case was higher than that
for the Gaussian case. We noted that the expected timescale also
varied. Between the Gaussian and the non-Gaussian case, we also
observed that the median value of the asymmetric distributions
decreases of roughly $20\%$. Concerning the expected number of
microlensing events, we estimated the same variation for sources
with a non-Gaussian VD, that is an increase of roughly $20\%$ from
the value of $\sim1.2$ events obtained with a Gaussian VD (Mancini
et al. \cite{mancini04}).
%
\subsection{MACHO Halo fraction and mass}
%
Following the methodology of Alcock et al. (\cite{alcock00}),
namely the method of maximum likelihood, we estimated the halo
fraction $f$ in the form of MACHOs and the most likely MACHO mass.
The likelihood function is
\begin{equation}
L\left(m,f\right)= \mathrm{exp}(-N_{\mathrm{exp}})
\prod_{i=1}^{N_{\mathrm{obs}}} \left[ E \,
{\cal{E}}(T_{\mathrm{E}_{i}}) \, \frac{\mathrm{d} \Gamma}
{\mathrm{d} T_{\mathrm{E}} } \left(T_{\mathrm{E}_{i}}, m \right)
\right]\, ,
\end{equation}
where $N_{\mathrm{exp}}$ is the total number of expected events,
and $\mathrm{d} \Gamma/\mathrm{d} T_{\mathrm{E}}$ is the sum of
the differential rates of the lens populations (MACHOs, LMC halo,
LMC disk+bar). The MACHO contribution is multiplied by $f$. The
product applies to the $N_{\mathrm{obs}}$ observed events. The
resulting likelihood contours are shown in Fig. \ref{Fig
Likelihood_Contours}, where the estimate of the differential rate
was performed using a Gaussian VD (solid line) and a universal VD
(dashed line). The probabilities were computed using a Bayesian
method with a prior uniform in $f$ and $m$. A spherical isothermal
distribution was used to describe the lens density in the MW and
LMC haloes.
\begin{figure}
\centering
\resizebox{8cm}{!}{\includegraphics{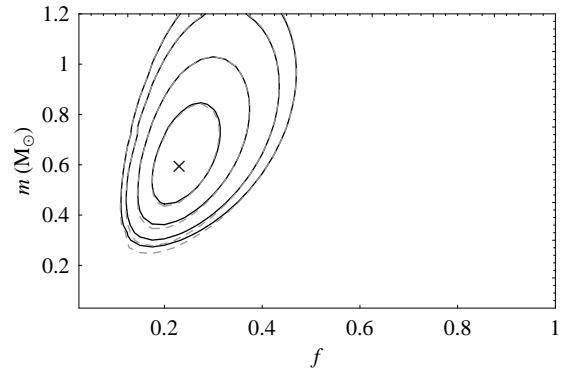}}
 \caption{Likelihood contours for MACHO mass $m$ and Halo fraction $f$
 for a typical spherical Halo. The contours enclose region of
 $34\%$, $68\%$, $90\%$, and $95\%$ probability.
 The cross shows the maximum-likelihood estimate.
 Two different curves are
 shown for each contour according to the velocity profile
 adopted for the sources: a Gaussian shape (solid line) and a
 universal VD (gray dotted line).}
 \label{Fig Likelihood_Contours}
\end{figure}
We found that the most probable mass for the Gaussian case is
$m=0.60_{-0.33}^{+0.40}$ M$_{\sun}$, where the errors are 68\%
confidence intervals, and $f=23\%$ with a $95\%$ confidence
interval of $10\%-47\%$. We note that these values are slightly
higher than, although fully compatible with, the original result
reported by Alcock et al. (\cite{alcock00}). The mismatch is due
to some differences in the modelling and the fact that the set of
the events considered is smaller. If we consider that the
velocities of the stars in the LMC are non-Gaussian-distributed,
the likelihood contours have minimal differences from those of the
previous case. We note that the most significant variation is in
the estimate of the lens mass, but that this is not higher than
$2\%$ for the $95\%$ probability contour line.

\section{Discussion and conclusion}
We have investigated the limits of the validity of the Gaussian
approximation used to describe the kinematics of a source
population in a microlensing context. This hypothesis, due to its
practicality, has always been adopted without any check of its
plausibility. We have remedied this deficiency in confirmation by
an exhaustive analysis. To describe the motion of a stellar
population with a non-spheroidal distribution as correctly as
possible, we utilize two VDs (Sect. 2.1, Sect. 2.2) extracted from
numerical simulations of collisionsless systems formed by
self-gravitating particles. These VDs are substantially different
than for a Gaussian one. We considered stars in the disk and bar
components of the LMC and investigated their potential to be
sources of lensing by transient lenses. In this framework, we
recalculated the main microlensing parameters, including the MACHO
halo fraction and the most likely value for the lens mass.

For a configuration in which the lenses and sources belong to the
target galaxy, we detected an increase in the differential rate of
microlensing events towards the LMC when we used a non-Gaussian VD
to describe the motion of its stars instead of a Gaussian one.
This increase is reflected in the estimate of the number of
expected events, which is roughly {\bf $20\%$} higher than the 1.2
events found for the Gaussian case.

The prediction for a halo that consists entirely of MACHOs is a
factor of $\sim5$ above the observed rates. The situation does not
change in a noticeable way if we consider a non-Gaussian VD, since
we have found that the number of events expected is practically
equal to the previous case. The results remain valid for both the
DT and the universal VD. The main difference between the velocity
dispersion of the LMC stars and the MACHOs, practically
neutralizes any possible variation due to the different shape of
the VD of the sources. The maximum-likelihood analysis provides
values for $m$ and $f$ that are quite similar for the Gaussian and
the non-Gaussian case. We conclude that the error in the estimate
of the most probable value for the MACHO mass as well as for the
Galactic halo fraction in the form of MACHOs, calculated with a
Gaussian VD for the LMC sources, is roughly of the order of
$1-2\%$. This fact implies that, in the study of the MW halo
composition by microlensing, a Gaussian profile is a reasonable
approximation for the velocity distribution of a system of source
stars, even if they are not spherically distributed. On the other
hand, for self-lensing, the Gaussian does not provide a good
description of the kinematics of a non-spherically distributed
stellar population, in a similar way to the disk or the bar of the
LMC. To ensure accurate microlensing predictions, it is thus
necessary to replace the Gaussian VD by a more physically
motivated one, which takes into account the real spatial
distribution of the source stars.

\begin{acknowledgements}
The author wish to thank Valerio Bozza, Gaetano Scarpetta, and the
anonymous referee for their contribute to improve the quality of
this work, and Steen Hansen and Sebastiano Calchi Novati for their
useful suggestions and communications. The author acknowledge
support for this work by funds of the Regione Campania, L.R.
n.5/2002, year 2005 (run by Gaetano Scarpetta), and by the Italian
Space Agency (ASI).

\end{acknowledgements}

------------------------------------------------------------------

\end{document}